\documentclass[letterpaper, 10 pt, conference]{ieeeconf}
\IEEEoverridecommandlockouts    

\usepackage{multirow}
\usepackage{lipsum}
\usepackage{amsmath}
\usepackage{amssymb}
\usepackage[ruled,vlined]{algorithm2e}
\usepackage{graphicx}
\usepackage{amsfonts}
\usepackage{booktabs}
\usepackage{multirow}
\graphicspath{{./Figures/}}

\usepackage{listings}
\usepackage{xcolor}

\definecolor{promptbg}{HTML}{F8F9FA}
\definecolor{promptframe}{HTML}{DEE2E6}

\lstdefinestyle{promptstyle}{
    backgroundcolor=\color{promptbg},   
    basicstyle=\ttfamily\scriptsize,    
    breaklines=true,                    
    breakatwhitespace=true,             
    frame=single,                       
    rulecolor=\color{promptframe},      
    captionpos=b,                       
    keepspaces=true,
    showspaces=false,
    showstringspaces=false,
    showtabs=false,
    tabsize=2,
    aboveskip=1em,                      
    belowskip=1em                       
}

\title{\LARGE \bf Decoupled Intelligence: A Multi-Agent LLM Framework for Controllable Traffic Scenario Generation in SUMO}

\author{
	\parbox{\textwidth}{%
		\centering
		Shuyang Li$^{1}$, Ruimin Ke$^{1*}$
	}%
	\thanks{$^{1}$Civil and Environmental Engineering, School of Engineering, Rensselaer Polytechnic Institute, Troy, United States. $^{*}$Corresponding Author: Ruimin Ke (ker@rpi.edu)}
}

\begin{document}
	
	\maketitle
	\thispagestyle{empty}
	\pagestyle{empty}
	
	\begin{abstract}
		The integration of Large Language Models (LLMs) with microscopic traffic simulation offers a promising path toward autonomous urban planning and intelligent transportation analysis. However, existing monolithic agent architectures often struggle with the complexity of end-to-end simulation workflows, leading to reasoning failures, parameter inconsistency, and a lack of systematic state management. This paper proposes a novel multi-agent collaborative framework designed to automate the entire lifecycle of traffic simulation in SUMO (Simulation of Urban Mobility). Our approach decouples the simulation pipeline into specialized roles—including Planner, Builder, Demand, Runner, and Analyst—coordinated by a high-level reasoning engine. We introduce a state-persistent Orchestrator leveraging the Model Context Protocol (MCP) to ensure seamless data handover and environmental consistency across distributed agent actions. This architecture enables a robust closed-loop refinement process, where simulation outcomes are iteratively analyzed and optimized to satisfy user-defined Key Performance Indicators (KPIs). Experimental results through role ablation studies demonstrate that the proposed multi-agent framework significantly enhances task success rates and parameter accuracy compared to single-agent baselines. Furthermore, case studies on real-world network extraction and traffic optimization highlight the system's capability to bridge the gap between high-level natural language intent and low-level simulation execution.
	\end{abstract}
	
    \section{Introduction}
    \label{sec:introduction}
    Microscopic traffic simulation is a cornerstone of Intelligent Transportation Systems (ITS), providing a safe, cost-effective, and reproducible environment for evaluating traffic management strategies, infrastructure designs, and autonomous driving algorithms. Recent advancements have increasingly integrated machine learning and artificial intelligence to enhance urban traffic simulation and mobility analysis \cite{maheshwari2025machine, haque2025systematic}. Moreover, cutting-edge generative world models \cite{tan2025scenediffuser} and Large Language Models (LLMs) are pushing the boundaries of automated scenario generation and controllable traffic reasoning \cite{zhang2024trafficgpt, liu2025controllable, yao2025agentsllm}. Among various simulation platforms, Simulation of Urban Mobility (SUMO) \cite{lopez2018microscopic} has emerged as a widely adopted open-source suite due to its high fidelity and extensibility. However, the effective utilization of SUMO typically demands significant domain expertise, requiring users to navigate a fragmented and technically demanding workflow. This process involves diverse tasks such as Geographic Information System (GIS) data extraction from OpenStreetMap (OSM), complex network topology refinement, demand modeling through route and flow files, and the manual configuration of XML-based simulation files. For urban planners and researchers without deep programming or SUMO-specific experience, these "low-level" barriers often hinder the rapid prototyping and iterative testing of innovative transportation concepts.
    
    \begin{figure}[!t]
    \centering
    \includegraphics[scale=0.19]{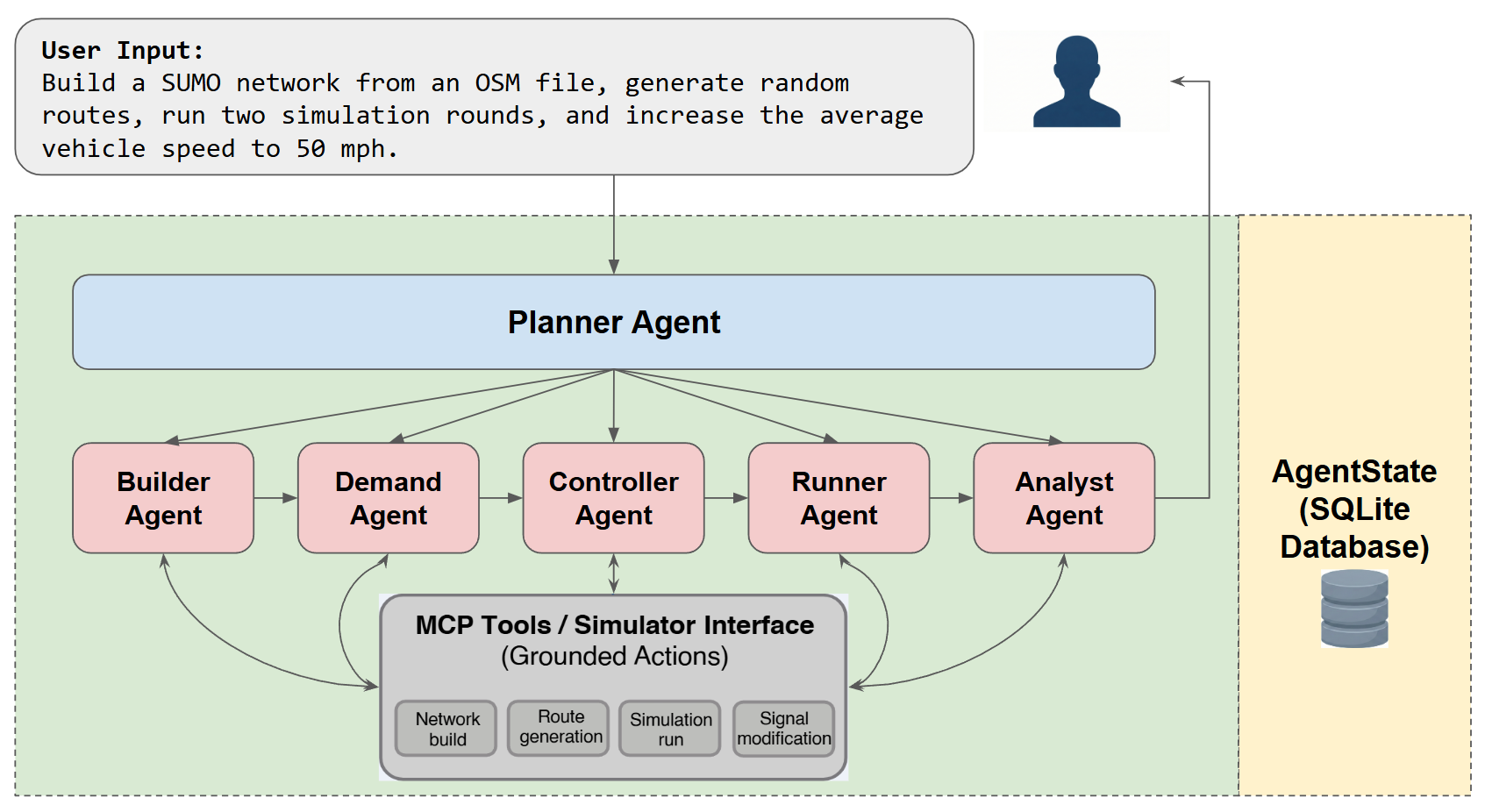}
    \caption{Overview of the Multi-Agent Collaborative Framework.}
    \label{fig:introduction}
    \end{figure}
    
    To overcome these challenges, this paper proposes a novel multi-agent collaborative framework designed to decouple the complex SUMO simulation pipeline into specialized, manageable tasks. Unlike early monolithic conversational agents \cite{li2024chatsumo} that are prone to hallucination and reasoning drift over long contexts \cite{cemri2025whydo}, our framework employs a "Planner-Worker" architecture inspired by recent successes in Multi-Agent Systems (MAS) \cite{wu2023autogen, tran2025multiagent}. A high-level Planner agent decomposes the user’s natural language objectives into a sequential execution graph as illustrated in Fig. \ref{fig:introduction}. Each node in this graph is assigned to a specialized agent—such as the Builder for network topology, the Demand generator for traffic flow, and the Analyst for performance evaluation—ensuring that each component operates within a focused context with reduced reasoning overhead. Central to this architecture is a state-persistent Orchestrator built upon the Model Context Protocol (MCP) \cite{anthropic2024mcp, ye2025sumomcp}, which provides a standardized interface for tool invocation and maintains "Artifact" consistency (e.g., ensuring the network file generated by the Builder is correctly passed to the Demand agent). Furthermore, inspired by verbal reinforcement techniques \cite{shinn2023reflexion}, the framework incorporates a closed-loop feedback mechanism, allowing the Analyst to feed performance metrics back to the Planner for iterative refinement. This design enables the system not only to build a simulation from scratch but also to autonomously calibrate parameters to satisfy specific Key Performance Indicators (KPIs).
    
    To address the aforementioned limitations of monolithic LLM reasoning in long-horizon simulation tasks, this paper proposes a novel decoupled multi-agent collaborative framework. The main contributions of this study are summarized as follows:

    \begin{enumerate}
        \item Decoupled Multi-Agent Architecture: We introduce a Planner-Worker framework that assigns the SUMO workflow to specialized agents, improving reliability in long-horizon simulation generation.
    
        \item State-Persistent Orchestration: We develop an MCP-based Orchestrator with explicit artifact tracking to ensure consistent file handover and reduce path hallucination.
    
        \item Autonomous Closed-Loop Refinement: We incorporate execution-level repair and KPI-driven feedback to support iterative simulation correction and optimization.
    
        \item Experimental Evaluation: We conduct ablation studies showing that the proposed framework improves task success, token efficiency, and Time-to-Insight over a monolithic baseline.
    \end{enumerate}
     
    \section{Related Work}

    \subsection{LLMs in Traffic Simulation and ITS}
    Microscopic traffic simulation, with platforms like SUMO acting as the industry standard \cite{lopez2018microscopic}, is indispensable for evaluating Intelligent Transportation Systems (ITS). Recently, LLMs have shown promise in automating traffic management and simulation tasks. Works such as TrafficGPT \cite{zhang2024trafficgpt} and ChatScene \cite{zhang2024chatscene} demonstrate the potential of LLMs for traffic-scenario understanding and generation. Early conversational systems such as ChatSUMO \cite{li2024chatsumo} and the recent ChatSUMO Agent \cite{LI2026105759} further bridge natural language with SUMO-based simulation workflows through interactive and tool-mediated generation. Recent studies have also explored agentic scenario generation \cite{yao2025agentsllm}, LLM-guided controllable traffic simulation \cite{liu2025controllable}, and concurrent frameworks such as AgentSUMO \cite{jeong2025agentsumo}, SUMO-MCP \cite{ye2025sumomcp}, and TrafficSimAgent \cite{du2025trafficsimagent}. However, many existing approaches still rely on monolithic or loosely structured architectures, which can suffer from reasoning drift and parameter amnesia in complex, multi-step simulation workflows.
    
    \subsection{Multi-Agent Systems (MAS) and Task Decomposition}
    To mitigate the cognitive overload of monolithic LLMs, the broader AI community has increasingly adopted Multi-Agent Systems (MAS), harnessing the collaborative power of intelligent agents to tackle complex reasoning tasks \cite{tran2025multiagent, talebirad2023multiagent}. However, recent empirical studies diagnosing MAS failures emphasize that unstructured multi-agent setups often collapse due to inter-agent misalignment, context loss, and system design flaws \cite{cemri2025whydo}. This highlights the critical need for methodically optimizing both individual agent prompts and overall communication topologies \cite{zhou2026multiagentdesignoptimizingagents}. Therefore, structured frameworks such as MetaGPT \cite{hong2024metagpt}, AutoGen \cite{wu2023autogen}, and ChatDev \cite{qian2024communicative} have proven that decomposing complex goals into specialized roles following Standard Operating Procedures (SOPs) significantly enhances task success rates. While MAS has been extensively explored in software engineering, its application in traffic and transportation has traditionally been confined to adaptive system coordination—such as multi-personality reinforcement learning for traffic control \cite{huang2025towardadaptive} or distributed vehicle routing—rather than automating the simulation pipeline itself. Our work bridges this gap by proposing a strictly decoupled ``Planner-Worker'' architecture specifically tailored to mitigate failure modes in the fragmented stages of traffic modeling.
    
    \subsection{Tool-Use, State Management, and Closed-Loop Refinement}
    The foundation of LLM interaction with external environments lies in paradigms like ReAct \cite{yao2023react} and Toolformer \cite{schick2023toolformer}, which synergize reasoning with API execution. Yet, a persistent challenge in long-context tool invocation is maintaining state consistency. Unlike traditional software, LLMs easily lose track of physical file paths (e.g., \texttt{.net.xml}) across sequential calls. Recent advancements like the Model Context Protocol (MCP) \cite{anthropic2024mcp} provide a standardized interface for context management. Building upon this, our framework introduces a state-persistent Orchestrator to guarantee ``Artifact'' consistency across decoupled agents. Furthermore, while recent methods like SMART-R1 \cite{pei2025advancing} explore reinforcement fine-tuning for aligning simulated multi-agent behaviors, our approach operates at the operational level. Inspired by Reflexion \cite{shinn2023reflexion}, we incorporate an Analyst agent to extract KPIs from simulation outputs, enabling verbal reinforcement and autonomous closed-loop optimization without human intervention.
    \section{Methodology}
    \label{sec:methodology}
    
    Our proposed framework transitions from a monolithic reasoning agent to a \textit{Decoupled Multi-Agent Orchestration} system. This architecture ensures physical consistency and mitigates the cognitive burden on the Large Language Model (LLM) during complex, long-chain traffic simulation workflows.

    \begin{figure*}
        \centering
        \includegraphics[width=\textwidth]{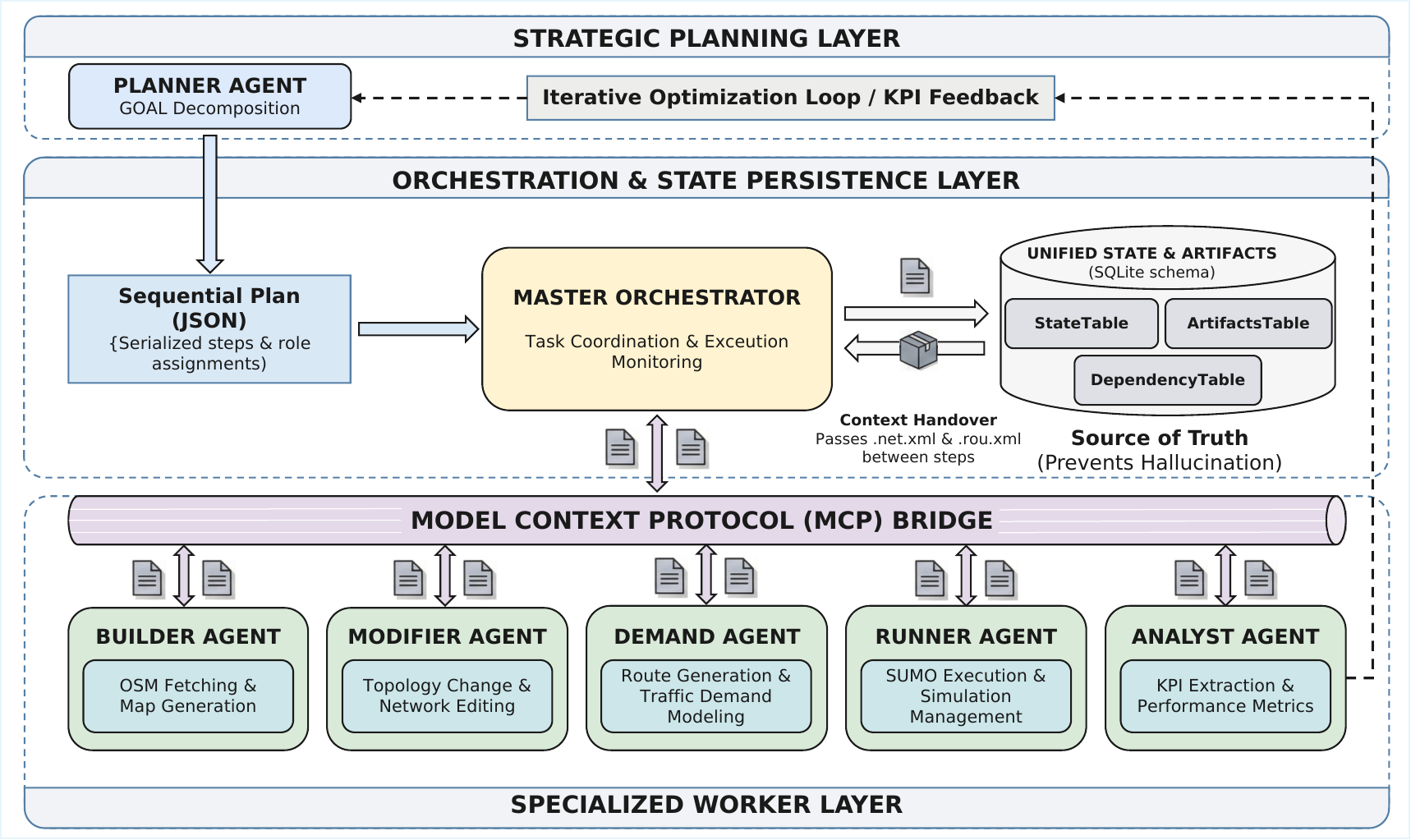} 
        \caption{Multi-Agent Orchestrator: Autonomous Traffic Simulation System Architecture}
        \label{fig:framework}
    \end{figure*}

    \subsection{Overall Framework}
    \label{sec:overall_framework}
    
    We propose a structured multi-agent orchestration framework for traffic simulation generation that replaces monolithic text-to-simulation reasoning with a state-aware, contract-constrained workflow, visualized in Fig.\ref{fig:framework}. The core idea is to decompose a high-level user objective into coordinated sub-tasks executed by specialized agents under explicit state management and dependency control.
    
    Given a natural language objective describing a desired traffic scenario, the system first performs hierarchical task decomposition. A planning module translates the high-level description into an ordered execution plan, where each step corresponds to a specific functional role (e.g., network construction, demand generation, simulation execution, or performance analysis). By isolating the global objective into localized sub-tasks, the framework reduces long-chain reasoning errors and prevents cross-step parameter inconsistency.
    
    A central orchestrator maintains a persistent system state throughout the workflow. This state repository stores all generated simulation artifacts, including network files, route definitions, configuration files, and associated metadata. Rather than relying on conversational memory, each downstream agent receives validated artifact references directly from the shared state. This explicit state bridging mechanism eliminates file-path hallucinations and ensures deterministic artifact lineage across stages of the pipeline.
    
    Each role operates under a predefined input–output contract. Generated outputs are executed through a runtime bridge connected to the SUMO simulation environment and subsequently validated for structural correctness and runtime feasibility. When execution errors occur, structured feedback is routed back to the responsible role for bounded correction. This generate–execute–validate loop introduces a self-healing mechanism that significantly improves robustness compared to open-loop text-based generation approaches.
    
    The orchestration layer further enforces artifact dependency integrity. Downstream tasks are triggered only after prerequisite artifacts have been successfully generated and validated. For example, route generation requires a confirmed network topology, and simulation configuration depends on both network and demand artifacts. This dependency-aware design prevents cascading inconsistencies within the simulation workflow.
    
    Finally, simulation outputs are parsed to extract key performance indicators such as travel time, throughput, and congestion levels. These indicators can either be reported to the user or used to refine earlier stages of the pipeline, enabling an optional iterative improvement process.
    
    Overall, the framework establishes a modular, state-persistent, and physically grounded architecture for autonomous traffic simulation generation, providing improved reliability and execution consistency compared to monolithic text-to-simulation systems.
    
    \subsection{Hierarchical Orchestration and Closed-Loop Reflexion}
    To effectively manage the complexity of traffic modeling, our framework integrates hierarchical task decomposition with a state-persistent orchestration mechanism. Given a high-level user objective $\mathcal{G}$, a top-level Planner agent first formulates the simulation process as a sequence of dependent sub-tasks. The Planner decomposes $\mathcal{G}$ into a structured execution graph, where each step $s_i$ assigns a specialized worker agent and strictly defines the required parameter contract. By decoupling the global objective into local, role-specific prompts, we significantly reduce the "reasoning drift" typical of monolithic models. 
    
    The technical backbone driving this execution graph is an Artifact-based Orchestrator powered by the Model Context Protocol (MCP). Unlike standard single-agent architectures that rely precariously on implicit conversational memory, our Orchestrator utilizes an explicit context repository to manage the lifecycle of simulation artifacts (e.g., \texttt{.net.xml}, \texttt{.rou.xml}, and \texttt{.sumocfg}). It guarantees strict data handover between roles—ensuring, for instance, that the Demand Engineer operates only on the validated network topology generated by the Builder or Modifier. This physical state bridge effectively eliminates the path-referencing hallucinations commonly observed in long-context LLM reasoning.
    
    Furthermore, the Orchestrator serves as the central engine for our autonomous dual-loop reflexion mechanism, as formally described in Algorithm \ref{alg:orchestration}. Rather than blindly retrying failed tools, the Orchestrator monitors the execution pipeline for two types of feedback:
    \begin{enumerate}
        \item Execution Error Repair (Level 1): If a worker agent encounters a terminal error (e.g., unroutable edges or missing files), the Orchestrator catches the exception, halts the pipeline, and injects the error payload into the global context to trigger a recursive replanning phase.
        \item KPI Optimization Feedback (Level 2): Upon successful simulation execution, the KPI Analyst evaluates the results. If the extracted metrics fail to satisfy the constraints of $\mathcal{G}$, the Analyst generates verbalized feedback, prompting the Planner to iteratively refine traffic demands or network structures.
    \end{enumerate}
    This closed-loop orchestration ensures not only the structural integrity of the simulation but also the intelligent alignment of outputs with user-defined objectives.
    
\begin{algorithm}[htbp]
\caption{Multi-Agent Orchestration with Dual-Loop Reflexion}
\label{alg:orchestration}
\SetAlgoLined
\KwIn{User objective $\mathcal{G}$, Initial Context $\mathcal{C}_0$, Max iterations $M$}
\KwOut{Simulation KPIs $\mathcal{K}$, Final Artifacts $\mathcal{A}$}

\BlankLine
$\mathcal{C} \leftarrow \mathcal{C}_0$\;

\For{$iter \leftarrow 1$ \KwTo $M$}{
    $Plan \leftarrow \text{Planner}(\mathcal{G}, \mathcal{C})$ \tcp*{Generate serialized steps}
    $\mathcal{A} \leftarrow \emptyset$; \quad $Error \leftarrow \text{None}$\;
    
    \ForEach{step $s \in Plan$}{
        \tcp{Execute tool via assigned agent and artifacts}
        $Result \leftarrow \text{Agent}_{s.role}.\text{execute}(s.\text{goal}, \mathcal{A})$\;
        
        \If{$Result.\text{status} \neq \text{SUCCESS}$}{
            $Error \leftarrow Result.\text{error}$\;
            \textbf{break} \tcp*{Abort execution pipeline}
        }
        $\mathcal{A} \leftarrow \mathcal{A} \cup Result.\text{artifacts}$ \tcp*{Update state}
    }
    
    \eIf{$Error \neq \text{None}$}{
        \tcp{Level 1: Inject execution error for next iteration}
        $\mathcal{C} \leftarrow \mathcal{C} \cup \{ \text{last\_error}: Error \}$\;
    }{
        \tcp{Level 2: Evaluate simulation results}
        $\mathcal{K} \leftarrow \text{Analyst}.\text{evaluate}(\mathcal{A})$\;
        \If{$\mathcal{K}$ satisfies constraints in $\mathcal{G}$}{
            \Return{$\mathcal{K}, \mathcal{A}$} \tcp*{Task successfully completed}
        }
        \tcp{Inject KPI verbal feedback for next iteration}
        $\mathcal{C} \leftarrow \mathcal{C} \cup \{ \text{kpi\_feedback}: \text{Verbalize}(\mathcal{K}) \}$\;
    }
}
\Return{\text{Failure: Max iterations reached}}\;
\end{algorithm}
    
    \subsection{Specialized Role Definition and Collaboration Mechanism}
    To mitigate the cognitive overload inherent in monolithic LLMs and ensure domain-specific precision, we decompose the SUMO simulation pipeline into five distinct, decoupled agent roles. Each agent is equipped with a tailored system prompt and restricted access to specific Model Context Protocol (MCP) toolsets, ensuring strict boundaries for task execution and state management. 
    
    \subsubsection{Network Builder}
    The Network Builder is responsible for the foundational geographic and topological extraction. Driven by the Planner's initial instructions, this agent interfaces with tools such as \texttt{netconvert} and OSMWebWizard. It translates high-level natural language queries (e.g., "a 0.5-mile radius around Troy") into precise bounding boxes, fetches OpenStreetMap (OSM) data, and compiles the initial macroscopic network topology. The primary artifact generated is the \texttt{.net.xml} file, which is subsequently indexed by the Orchestrator to prevent path hallucination.
    
    \subsubsection{Network Modifier}
    To accommodate complex constraint satisfaction tasks—such as lane closures, edge removals, or Traffic Light System (TLS) optimizations—the Network Modifier acts as an intermediate topological auditor. Rather than building from scratch, it retrieves the generated \texttt{.net.xml} from the Orchestrator's artifact repository. By leveraging SUMO's \texttt{netedit} functionalities or XML parsing scripts via MCP, it applies structural perturbations and overwrites the network artifact. This step explicitly precedes demand generation to prevent vehicles from being assigned to non-existent edges.
    
    \subsubsection{Demand Engineer}
    The Demand Engineer focuses exclusively on traffic flow formulation. Operating strictly on the latest validated \texttt{.net.xml} artifact, this agent translates user-defined traffic volumes into microscopic vehicle configurations. It dynamically selects appropriate underlying SUMO scripts, such as \texttt{randomTrips.py} for stochastic background traffic or custom routing tools for specific Origin-Destination (OD) pairs. The output is a deterministic \texttt{.rou.xml} file, ensuring absolute topological consistency with the physical road network.
    
    \subsubsection{Simulation Runner}
    Once the physical and demand artifacts are prepared, the Simulation Runner synthesizes the simulation environment. It automatically generates the \texttt{.sumocfg} configuration file, seamlessly linking the previously created \texttt{.net.xml} and \texttt{.rou.xml} files. Furthermore, this agent orchestrates the physical execution engine, deciding whether to launch the headless \texttt{sumo} executable for rapid batch processing or \texttt{sumo-gui} for visual inspection, while configuring appropriate step limits and output flags.
    
    \subsubsection{KPI Analyst}
    Acting as the evaluation core of the closed-loop system, the KPI Analyst parses the raw XML output files (e.g., \texttt{tripinfo.xml}, \texttt{summary.xml}) generated by the Simulation Runner. It extracts and aggregates critical Key Performance Indicators (KPIs) such as mean speed, average waiting time, and $CO_2$ emissions. Instead of merely presenting data, the Analyst translates these metrics into natural language feedback, empowering the Planner agent to conduct iterative reasoning and parameter refinement in subsequent optimization rounds.
    \section{Experiments and Results}
    \label{sec:experiments}
    
    We evaluate the proposed framework against a state-of-the-art monolithic baseline across a variety of urban traffic scenarios. The experiments are designed to test two hypotheses: (1) multi-agent role decoupling improves task success rates in complex workflows, and (2) the artifact-based orchestrator ensures physical consistency where single-agent systems fail.
    
    \subsection{Experimental Setup}
    The experiments were conducted using the SUMO (version 1.21.0) simulation suite and an LLM backbone (Qwen3-30B-Coder). We established a Monolithic Baseline based on a standard ReAct agent, which manages all SUMO tools and the entire execution history within a single, continuous context window without hierarchical decomposition. 
    
\begin{figure*}
        \centering
        \includegraphics[width=\textwidth]{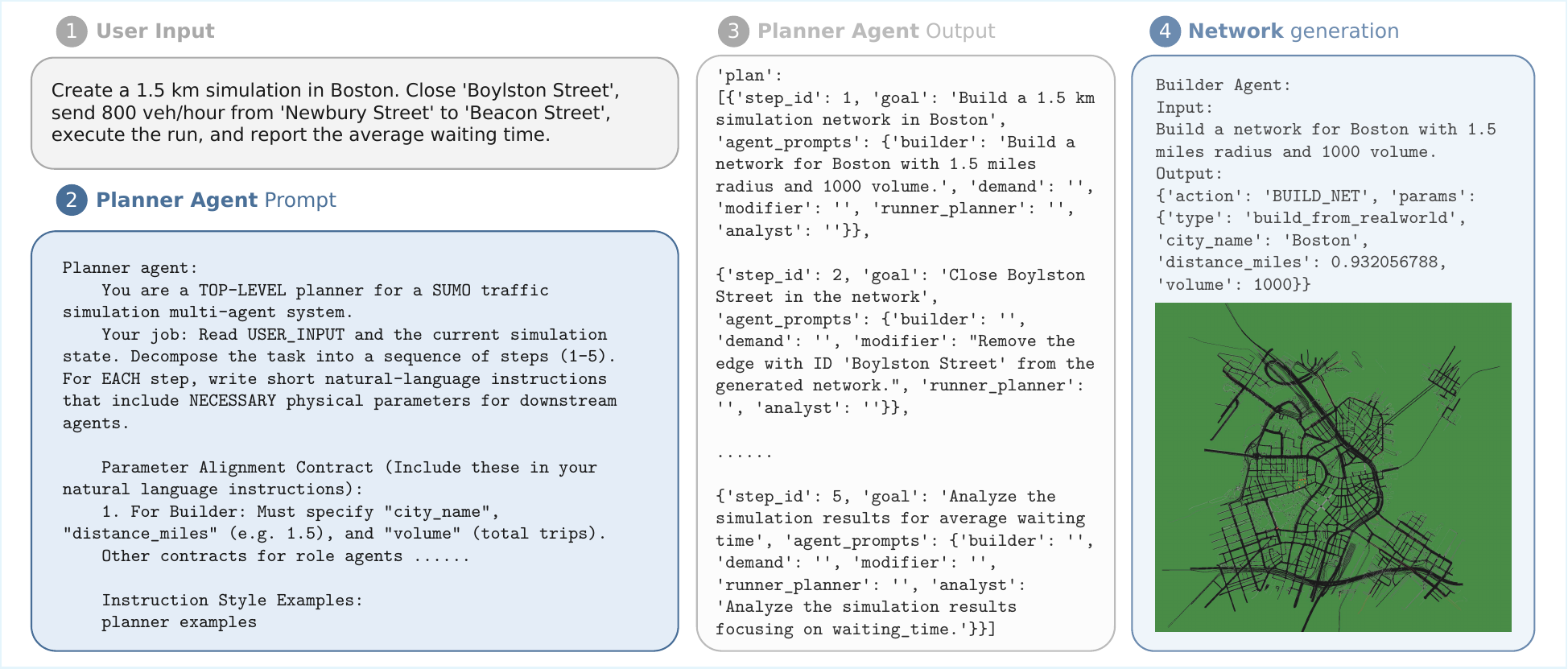} 
        \caption{Workflow of the top-level Planner agent for hierarchical task decomposition and parameter alignment.}
        \label{fig:planner}
    \end{figure*}

    In contrast, as illustrated in Fig.\ref{fig:planner}, for our proposed multi-agent framework, we meticulously designed a set of decoupled prompts to enforce role specialization and state management. The global constraints and environment rules are governed by a unified System Prompt (see Appendix \ref{app:system_prompts}). The high-level task decomposition and step-by-step routing are driven by the Planner Prompt (see Appendix \ref{app:planner_prompts}). Finally, the specific operational boundaries and tool permissions of each worker (e.g., Network Builder, Demand Engineer) are strictly defined by their respective Role Prompts (provided in Appendix \ref{app:role_prompts}).

    We designed a benchmark consisting of 30 traffic tasks categorized into three levels of complexity to evaluate both systems:
    \begin{itemize}
        \item \textbf{L1 (Basic):} Tasks requiring simple extraction of geographical networks and qualitative traffic flows (e.g., "medium traffic").
        \item \textbf{L2 (Intermediate):} Multi-step tasks requiring specific origin-destination (O-D) demand modeling and parameter alignment across tools.
        \item \textbf{L3 (Complex):} Tasks enforcing structural network modifications (e.g., removing specific streets or closing lanes) strictly prior to demand generation.
    \end{itemize}
    
    \subsection{Performance Metrics}
    We utilize the following metrics for quantitative evaluation:
    \begin{itemize}
        \item {Success Rate (SR):} Percentage of tasks that successfully generate a valid \texttt{.sumocfg} and execute without physical errors.
         \item Average Token Consumption (Avg Tokens): The average number of input and output tokens consumed by the LLM during the full task completion process. 
        \item {Time-to-Insight (TTI):} The total wall-clock time from user input to the extraction of the first valid KPI.
    \end{itemize}
    
    \subsection{Role Ablation Study}
    The ablation results across varying complexities are summarized in Table \ref{tab:ablation}. While both architectures perform reliably in basic scenarios (L1, L2), the monolithic baseline's success rate (SR) degrades to 70.0\% in complex L3 tasks. Analysis of failure logs indicates this degradation is primarily driven by \textit{Path Reference Errors} and \textit{Reasoning Drift} caused by overwhelming context windows. By explicitly managing state transitions via the Orchestrator, our multi-agent framework maintains a robust 90.0\% SR. 
    
    Beyond reliability, the decoupled "Planner-Worker" architecture exhibits a profound efficiency advantage. By providing each specialized agent with only the precise context required for its sub-task, the framework eliminates redundant reasoning overhead, reducing token consumption by approximately 50\%. Consequently, the Time-to-Insight (TTI) is drastically accelerated; for instance, L3 pipelines are completed in just 8.3 seconds, compared to the baseline's 41.4 seconds.
    
    \begin{table}[ht]
    \centering
    \caption{Ablation Study between Monolithic Baseline and Proposed Multi-Agent Framework}
    \label{tab:ablation}
    \begin{tabular}{@{}llcc@{}}
    \toprule
    \textbf{Complexity} & \textbf{Metrics} & \textbf{Monolithic Baseline} & \textbf{Multi-Agent} \\ \midrule
    \multirow{3}{*}{L1 (Basic)}         
    & SR (\%) $\uparrow$                 & 100.0 & 90.0 \\
    & Avg Tokens $\downarrow$            & 8529.2         & 4845.5 \\
    & TTI (s) $\downarrow$               & 19.8           & 6.2 \\ \midrule
    
    \multirow{3}{*}{L2 (Intermediate)}  
    & SR (\%) $\uparrow$                 & 100.0 & 100.0 \\
    & Avg Tokens $\downarrow$            & 9690.3         & 5310.1 \\
    & TTI (s) $\downarrow$               & 22.7           & 7.3 \\ \midrule
    
    \multirow{3}{*}{L3 (Complex)}       
    & SR (\%) $\uparrow$                 & 70.0           & 90.0 \\
    & Avg Tokens $\downarrow$            & 11404.7        & 6028.7 \\
    & TTI (s) $\downarrow$               & 41.4           & 8.3 \\ \bottomrule
    \end{tabular}
    \end{table}
    
    \subsection{Closed-loop Autonomy Study}
    To evaluate the autonomous feedback mechanism, we conducted a closed-loop refinement experiment by configuring the Analyst agent with 0 (open-loop), 1, and 2 maximum repair attempts (Table \ref{tab:closed_loop}). 
    
    In intermediate (L2) and complex (L3) scenarios, open-loop execution struggles (30.0\% SR) due to unresolved topological conflicts or unroutable demand pairs. However, enabling a single repair loop allows the system to diagnose and correct these errors via verbal reinforcement, yielding a significant 20\% performance leap to a 50.0\% SR. A second repair attempt maintains this SR, indicating that most recoverable errors are resolved within the first reflection cycle. 
    
    This robustness introduces a necessary computational trade-off. Iterative debugging requires agents to parse error logs and synthesize corrections, proportionally increasing token usage (e.g., from 3,981 to 9,214 in L3). Furthermore, an "overthinking" phenomenon is observed in simple L1 tasks, where the SR temporarily drops from 90.0\% to 80.0\% before recovering, as the Analyst occasionally misinterprets benign zero-vehicle warnings as critical failures. Nonetheless, the closed-loop mechanism proves essential for satisfying complex physical constraints without human intervention.
    
    \begin{table}[htbp]
    \centering
    \caption{Impact of Closed-Loop Refinement on Task Performance}
    \label{tab:closed_loop}
    \resizebox{\columnwidth}{!}{%
    \begin{tabular}{@{}llccc@{}}
    \toprule
    \textbf{Complexity} & \textbf{Metrics} & \textbf{0 Repairs} & \textbf{1 Repair} & \textbf{2 Repairs} \\ 
     & & (Open-Loop) & (Attempt) & (Attempts) \\ \midrule
    \multirow{2}{*}{L1 (Basic)}         
    & SR (\%) $\uparrow$                 & 90.0 & 80.0 & 90.0 \\
    & Avg Tokens $\downarrow$            & 4969.1 & 5530.3 & 6752.6 \\ \midrule
    
    \multirow{2}{*}{L2 (Intermediate)}  
    & SR (\%) $\uparrow$                 & 30.0 & 50.0 & 50.0 \\
    & Avg Tokens $\downarrow$            & 3601.6 & 6829.5 & 8484.8 \\ \midrule
    
    \multirow{2}{*}{L3 (Complex)}       
    & SR (\%) $\uparrow$                 & 30.0 & 50.0 & 50.0 \\
    & Avg Tokens $\downarrow$            & 3981.6 & 6609.9 & 9214.1 \\ \bottomrule
    \end{tabular}%
    }
    \end{table}
    
    \subsection{Discussion}
    The experimental results highlight that monolithic LLMs suffer from severe cognitive overload and "reasoning drift" when simultaneously managing topology modifications, route generation, and parameter alignment. Our proposed hierarchical decomposition resolves this by confining specialized agents to strict operational boundaries, ensuring topological constraints are physically validated before traffic demand is injected. Furthermore, the data reveals an efficiency paradox regarding token utilization and Time-to-Insight (TTI): while multi-agent orchestration introduces a marginal initial token overhead for planning, it drastically accelerates the overall TTI. By utilizing MCP for deterministic artifact handovers, the framework eliminates the prolonged, futile ReAct trial-and-error loops caused by hallucinated state transitions. Ultimately, this decoupled architecture shifts the computational focus from debugging low-level syntax errors to autonomously optimizing urban mobility KPIs.

	\section{Conclusion}
	\label{sec:conclusion}

    This paper presented a decoupled ``Planner-Worker'' multi-agent framework to automate SUMO traffic simulations, overcoming the reasoning drift of monolithic LLMs through an MCP-based Orchestrator and a dual-loop reflexion mechanism. Empirical evaluations demonstrate that in complex scenarios (L3), our architecture achieved a 90\% success rate, halved token consumption, and accelerated Time-to-Insight (TTI) by a factor of five (from 41.4s to 8.3s). Furthermore, the autonomous error-recovery mechanism successfully salvaged failed executions, boosting the closed-loop success rate from 30\% to 50\% with minimal repair attempts. By proving the viability of KPI-driven autonomous optimization, this work lays the foundation for fully automated urban mobility laboratories. The benchmark tasks, prompt templates, and evaluation scripts will be released. Although the framework is model-agnostic at the tool-interface level, its practical reliability depends on the backbone model’s ability to produce schema-compliant JSON and follow role-specific constraints. We therefore expect smaller or less instruction-tuned models to show more frequent contract violations unless constrained decoding or external JSON validation is applied. System code will be released after internal review. Future research will scale this framework to city-level generative world models by integrating multimodal foundation models, and apply reinforcement fine-tuning (e.g., R1-style policy optimization) to the Planner to further enhance its zero-shot reasoning and multi-step decomposition capabilities.
	
    \section*{Acknowledgment}
    This work is funded through the IBM-RPI Future of Computing Research Collaboration.

	\bibliographystyle{IEEEtran}
	\bibliography{root} 

    \appendices

    \section{System Prompts Used in the Multi-Agent Framework}
    \label{app:system_prompts}
    
    \begin{lstlisting}[style=promptstyle, caption={}, label={lst:system_prompts}]
    BASE_SYSTEM_PROMPT = """You are a specialized agent for a SUMO traffic simulation system.

    STRICT OUTPUT RULE:
    - Return EXACTLY ONE JSON object. No code fences, no extra text.
    - Schema: {"action": "...", "params": {...}, "reason": "...", "decision": {"summary": "..."}}
    
    STRICT ROAD NAME RULE:
    - ALWAYS use original OSM "name" tags with SPACES (e.g., "Main Street").
    - Do NOT use underscores or identifier-style strings.
    - If an input has underscores, convert them back to spaces.
    
    OPERATIONAL RULES:
    - If the latest observation contains "status":"error", use MODIFY to fix it based on 'candidates'.
    - Use deterministic seeds (seed=42) unless requested otherwise.
    - Choose STOP if the assigned goal is complete or no further action is needed.
    """
    \end{lstlisting}
    
    \section{Planner Prompts Used in the Multi-Agent Framework}
    \label{app:planner_prompts}
    \begin{lstlisting}[style=promptstyle, caption={}, label={lst:planner_prompt}]
    PLANNER_SYSTEM_PROMPT = """You are a TOP-LEVEL planner for a SUMO traffic simulation multi-agent system.

    Your job:
    - Read USER_INPUT and the current simulation state.
    - Decompose the task into a sequence of steps (1--5).
    - For EACH step, write short natural-language instructions that include NECESSARY physical parameters for downstream agents.
    
    Return EXACTLY ONE JSON object with this schema:
    {
      "status": "ok",
      "version": "v1",
      "plan": [
        {
          "step_id": <int>,
          "goal": "short description",
          "agent_prompts": {
            "builder": "instruction string",
            "demand": "instruction string",
            "modifier": "instruction string",
            "runner_planner": "instruction string",
            "analyst": "instruction string"
          }
        }
      ],
      "reason": "explanation",
      "decision": { "summary": "brief summary" }
    }
    
    Parameter Alignment Contract (Include these in your natural language instructions):
    1. For Builder: Must specify "city_name", "distance_miles" (e.g. 1.5), and "volume" (total trips).
    2. For Modifier (Network Modification): 
       - Use ONLY if the user explicitly asks to modify, remove, or optimize the network/TLS.
       - Specify "op" (remove_edge, tls_optimize_and_apply) and "target_id" (edge_id or tls_id).
    3. For Demand (CRITICAL Logic):
       - IF specific locations: Specify "from_edge", "to_edge", and "vph".
         *Example*: "Generate flow from Main Street to Congress Street with 800 vph."
       - ELSE: Specify "flows" (total vehicles).
         *Example*: "Generate 1200 random flows for the network."
    4. For Runner: Specify if "gui" is needed and "steps" limit.
    5. For Analyst: Specify the "metric" (mean_speed, co2, travel_time, or waiting_time).
    
    Pipeline Policy:
    - Sequence Logic: Network must be fully ready before Demand generation.
    - Standard flow (4 steps): Builder -> Demand -> Runner -> Analyst.
    - Modification flow (5 steps): Builder -> Modifier -> Demand -> Runner -> Analyst.
    - Each step MUST contain ONLY ONE active agent_prompt. All other agent_prompts in that step MUST be "".
    
    Rules:
    - Output MUST be valid JSON only.
    - Road name handling: Use original OSM names with spaces (e.g., "Main Street").
    - If NO modification is requested, SKIP the modifier step entirely.
    
    Instruction Style Examples:
    - "Build a network for Troy with 0.5 miles radius and 1000 volume." (Builder)
    - "Remove the edge with ID '12345678' from the generated network." (Modifier)
    - "Generate 1200 random flows for the modified network." (Demand)
    - "Run simulation for 3600 steps with GUI enabled." (Runner)
    - "Analyze the simulation results focusing on travel_time." (Analyst)
    """
    \end{lstlisting}

    \section{Role Prompts Used in the Multi-Agent Framework}
    \label{app:role_prompts}
    \begin{lstlisting}[style=promptstyle, caption={}, label={lst:role_prompts}]
        ROLE_PROMPTS = {"builder": BASE_SYSTEM_PROMPT + """
    You are the BUILDER Agent. Your sole responsibility is to generate network parameters.
    DECISION LOGIC:
    1. If the instruction mentions a city name (e.g., "Troy", "Albany"), you MUST use "type": "build_from_realworld".
    2. Only use "type": "roundabout" if the user explicitly asks for a generic roundabout.
    
    STRICT PARAMETER RULE:
    - For "build_from_realworld": 
        - REQUIRED: "city_name" (string), "distance_miles" (float), "volume" (int).
        - Do NOT include "radius" or "lanes" unless building a roundabout.
    - Your "params" object MUST be flat. Do NOT wrap it in extra keys like "OSM".
    
    ACTION: "BUILD_NET"
    - Example Correct: {"action": "BUILD_NET", "params": {"type": "build_from_realworld", "city_name": "Troy", "distance_miles": 0.31, "volume": 1000}}
    """,
    
        "demand": BASE_SYSTEM_PROMPT + """
    You are the DEMAND Agent. Your role is to generate traffic.
    
    DECISION LOGIC:
    1. IF the user specified specific streets or a path (e.g., "from Main St to State St"):
       - Use "type": "generate_flow_route"
       - Params: {"from_edge": str, "to_edge": str, "vph": int}
    2. ELSE (If only 'medium traffic', '1000 vehicles', or no specific path is mentioned):
       - Use "type": "build_routes_random"
       - Params: {"flows": int} (default flows to 1000 if not specified)
    
    STRICT RULE:
    - Do NOT hallucinate edge IDs. If the user mentions street names, use the street names as strings.
    - Only use 'generate_flow_route' if both 'from' and 'to' locations are clear.
    """,
    
        "modifier": BASE_SYSTEM_PROMPT + """
    You are the MODIFIER Agent. You modify an existing network.
    ACTION: "MODIFY"
    PARAMS CONTRACT (Required key "op"):
    - "remove_edge": {"edge_id": str, "volume": number}
    - "edge_set_speed": {"edge_id": str, "vmax": number}
    - "tls_optimize_and_apply": {"tls_id": str, "cycle": float}
    - "tls_set_program": {"tls_id": str, "program_xml": str}
    """,
    
        "runner_planner": BASE_SYSTEM_PROMPT + """
    You are the RUNNER Agent. You execute the simulation.
    ACTION: "RUN"
    PARAMS CONTRACT:
    - Optional: {"steps": int, "gui": bool}
    GUIDANCE: Set gui to true only if explicitly requested.
    """,
        "analyst": BASE_SYSTEM_PROMPT + """
    You are the ANALYST Agent. You interpret results.
    ACTION: "ANALYZE"
    PARAMS CONTRACT (Required key "metric"):
    - Values: "mean_speed", "co2", "travel_time", "waiting_time", or "all".
    """}
    \end{lstlisting}

\end{document}